\documentclass{aa}

\usepackage{graphicx}
\usepackage{txfonts}
\usepackage{amsmath}
\usepackage{placeins}
\usepackage{xcolor}
\usepackage{longtable}

\usepackage[colorlinks=true,linkcolor=blue,citecolor=blue,urlcolor=blue]{hyperref}
\usepackage{tabularx}
\usepackage{subcaption}
\usepackage{lscape}
\usepackage{lineno}

\title{Multi-scale variability in the optical afterglow of GRB~251013C from high-cadence LAST observations}

\titlerunning{LAST GRB~251013C}
\authorrunning{R. Konno et al.}

\author{
R.~Konno\inst{1}
\and E.~O.~Ofek\inst{1}
\and S.~Garrappa\inst{1}
\and O.~Teboul\inst{1}
\and E.~Waxman\inst{1}
\and S.~Ben-Ami\inst{1}
\and D.~Kovaleva\inst{1}
\and A.~Krassilchtchikov\inst{1}
\and D.~Polishook\inst{1}
\and E.~Segre\inst{1}
\and E.~A.~Zimmerman\inst{1}
}

\institute{
Department of Particle Physics and Astrophysics, Weizmann Institute of Science, 76100 Rehovot, Israel
}

\begin{document}
\nolinenumbers
\abstract
{Optical afterglows of gamma-ray bursts (GRBs) are often described by smooth power-law decays, but deviations such as flares and rebrightenings provide important constraints on the underlying emission processes.}
{We investigate the temporal variability of the optical afterglow of GRB~251013C using high-cadence observations, and assess the origin of the observed variability.}
{We analyze continuous optical observations obtained with the Large Array Survey Telescope (LAST), complemented by publicly available X-ray data from \textit{Swift}-XRT. The optical light curve is modeled using a power-law decay with superposed variability components, and contemporaneous optical and X-ray measurements are used to constrain the broadband spectrum.}
{The optical light curve shows pronounced variability on multiple timescales, including broad rebrightening episodes with $\Delta t \sim t$ and superposed faster fluctuations with $\Delta t / t < 1$. The main rebrightening exhibits a structured rise with a clear steepening prior to the peak. The fast variability comprises asymmetric fast-rise, slow-decay features whose durations increase with peak time. The optical flux lies on the extrapolation of the X-ray spectrum, and the spectrum hardens by the same amount whether measured within the X-ray band or from optical to X-ray.}
{We favor a refreshed external-shock interpretation for the main rebrightening. The broadband spectrum and post-peak decay are consistent with slow-cooling synchrotron emission in a wind-like medium. The spectral hardening requires a newly dominant, harder electron population, while the faster optical variability indicates structure within the refreshed ejecta or shocked region.}
{}

\maketitle

\section{Introduction}
\label{sec:Introduction}

Gamma-ray bursts (GRBs) are the most luminous explosions known, releasing up to $\sim10^{54}$~erg of isotropic equivalent energy within seconds \citep{Klebesadel1973, Kouveliotou1993}. GRBs are commonly divided into two broad classes based on their duration, short and long GRBs. Long-duration GRBs are associated with the collapse of massive stars \citep{WoosleyBloom2006}, and their afterglows provide crucial insight into the physics of relativistic jets and their interaction with the surrounding medium \citep{MeszarosRees1997, Sari1998, ZhangMeszaros2004}. 

Optical afterglows typically exhibit a smooth power-law (PL) decay in time, consistent with synchrotron radiation from an external forward shock \citep{Waxman1997, Sari1998, WijersGalama1999}. However, deviations from a simple monotonic decay are frequently observed, including flares, bumps, and short-timescale variability \citep{PanaitescuKumar2000, Bersier2003, Lipkin2004, Greiner2009}. These features may arise from refreshed shocks, density inhomogeneities in the circumburst medium, angular jet structure, or late-time central engine activity \citep{ReesMeszaros1998, Granot2003, Nakar2003, Zhang2006}. High-cadence optical monitoring is essential for resolving such rapid temporal variations, particularly within the first hours after the burst when the afterglow is brightest. 

The Large Array Survey Telescope (LAST) is a wide-field optical survey instrument optimized for high-cadence transient observations \citep{Ofek2023, BenAmi2023}. At the time of these observations it comprised 40 telescopes (28 cm, f/$2.2$), each providing a $\cong 7.4~$deg$^{2}$ field of view with a $1.25''\,$pixel$^{-1}$ sampling. Continuous photometric sequences obtained by LAST enable the study of afterglow variability on timescales of seconds to minutes, offering new constraints on the dynamics of the forward shock and potential signatures of internal or reverse-shock emission components.

In this work, we present LAST observations of the optical afterglow of the long GRB 251013C, focusing on the detection and characterization of photometric variability during the early decay phase. We describe the observations and data reduction in Sect.~\ref{sec:data}, present the temporal analysis in Sect.~\ref{sec:analysis}, provide physical implications in Sect.~\ref{sec:interpretation}, discuss the results in Sect.~\ref{sec:discussion}, and summarize our conclusions in Sect.~\ref{sec:conclusion}.

\begin{figure*}[t]
\centerline{\includegraphics[width=0.9\textwidth]{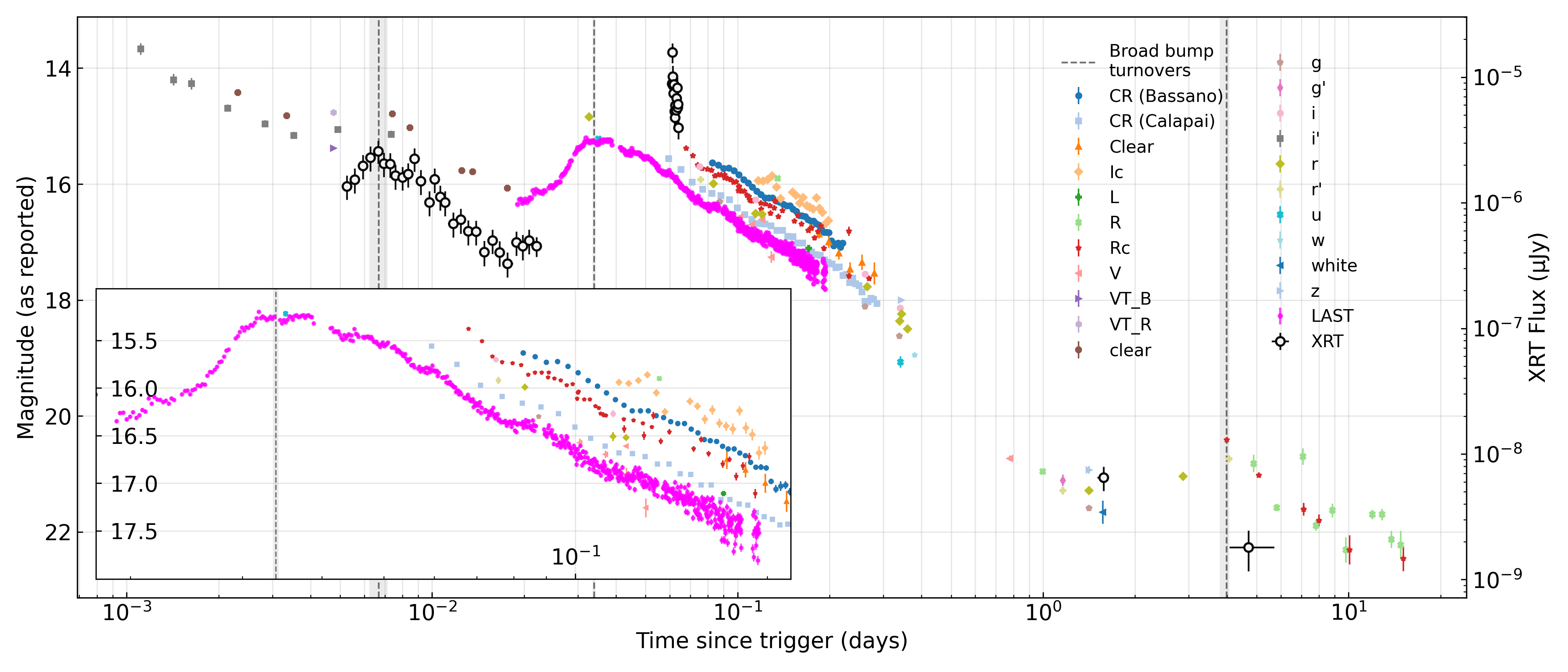}}
    \caption{Composite optical, UV and X-ray light curve of GRB~251013C. Magnitudes are shown as reported within the references and plotted versus time since $T_0 =$ 2025 October 13 17:39:41 UT. All optical--UV data except for LAST has been compiled from General Coordinates Network (GCN) circulars. Legend labels correspond to the following GCN sources; $g/g'$: \citep{GCN42229,GCN42276,GCN42279,GCN42334,GCN42569}, $r/r'/R/R_c$: \citep{GCN42227,GCN42276,GCN42279,GCN42228,GCN42229,GCN42230,GCN42242,GCN42253, GCN42254,GCN42569,GCN42266,GCN42267,GCN42269, GCN42283, GCN42301,GCN42319,GCN42333,GCN42334,GCN42338, GCN42399,GCN42736}, $i/i'/I_c$: \citep{GCN42229,GCN42256,GCN42333,GCN42334,GCN42253,GCN42569}, $z$: \citep{GCN42279,GCN42334}, $u$/\textit{white}: \citep{GCN42298,GCN42334}, $V$: \citep{GCN42261,GCN42269}, $w$: \citep{GCN42241}, $L$: \citep{GCN42231}, VT$_B$, VT$_R$: \citep{GCN42223}, CR (Bassano): \citep{GCN42262}, CR (Calapai): \citep{GCN42275}, clear/Clear: \citep{GCN42225, GCN42396}. \textit{Swift}-XRT data are taken from the online \textit{Swift}-XRT GRB Catalogue \citep{Evans2009}. The turnover times for three broad optical bumps are marked by dashed vertical lines.
}
\label{fig:GRB251013C_lightcurve_with_inset}
\end{figure*}

\section{Data}
\label{sec:data}

GRB~251013C was first detected on 2025 October 13 by the \textit{Fermi} Gamma-ray Burst Monitor (\textit{Fermi}-GBM; \citealt{Meegan2009}) at 17:39:41~UTC \citep{GBM_GCN42221}, and independently by ECLAIRs \citep{Cordier2015} on board the Space-based multi-band astronomical Variable Objects Monitor (SVOM) satellite at 17:39:42~UTC \citep{ECLAIRs_GCN42222}.

The SVOM-ECLAIRs alert was received by LAST at 17:40:51~UTC, while the \textit{Fermi}-GBM alert was received at 17:49:06~UTC. Both alerts were processed by the fully automated LAST target-of-opportunity (ToO) handler. The ToO handler evaluates incoming alerts according to predefined scientific and observational criteria and either accepts or rejects them. For accepted alerts, an observation plan is generated and forwarded to the scheduler, which executes observations based on target visibility, priority, and completion status. 

LAST observations commenced at 18:06:43~UTC, corresponding to $T-T_0 = 1622$~s, where $T$ denotes the start time of LAST observations and $T_0$ is the \textit{Fermi}-GBM trigger time \citep{GBM_GCN42221}. The data for GRB~251013C were reduced and filtered in real time by the LAST photometric and transients pipelines \citep{Ofek2023PipeI, Konno2026}, enabling rapid reporting of the optical transient to the internal transient marshal and a manual extension of observations. As a result, the GRB was monitored for $\sim4.2$~h during the first night, until it no longer satisfied observability constraints.

The afterglow was clearly detected throughout the first night, reaching a peak brightness of $m_\mathrm{LAST}\sim15.2$~mag. The magnitude $m_\mathrm{LAST}$ is given in the native unfiltered LAST AB system. The standard LAST observation strategy consists of taking $20\times20~$s exposures defined as visits, which are coadded to form a deeper image. The optical counterpart of the GRB was, however, detectable in individual $20$~s exposures, allowing construction of a light curve with $20$~s temporal resolution. 

Nominal LAST photometry results are derived with aperture radii of $2$, $4$, and $6$ pixels, as well as a PSF-fitting procedure. The single-exposure light curves obtained with aperture and PSF-fitting photometry agree well when the source is at its brightest. However, PSF photometry yields higher precision as the afterglow fades and therefore we adopt the PSF-fitting photometry for the remainder of this work. Nearby stars recorded in the same images and reduced identically over the same interval do not show similar fluctuations, although their measured scatter indicates an underestimation of the photometric uncertainties (App.~\ref{app:field_stars}). We use the nearby star population to estimate a time-dependent correction factor for the photometric uncertainties of GRB~251013C (App.~\ref{app:errcal}) and employ it in the analyses within this work. Given the consistency between independent photometric methods and external light curves (Fig.~\ref{fig:GRB251013C_lightcurve_with_inset}), we conclude that the observed fluctuations are intrinsic to the source.

The light curve of GRB~251013C as drawn from LAST data and publicly reported measurements (Fig.~\ref{fig:GRB251013C_lightcurve_with_inset}) exhibits pronounced afterglow variability from seconds to days after the burst. Up to three distinct rebrightening episodes are observed, with durations comparable to their corresponding observer times, in addition to shorter-timescale fluctuations resolved during the LAST observing window. This wide range of temporal variability clearly deviates from a canonical PL afterglow decay and provides a useful laboratory for studying variability mechanisms in GRB afterglows.

\section{Analysis}
\label{sec:analysis}

\subsection{Light curve preparation}

\begin{figure}
\centerline{\includegraphics[width=0.5\textwidth]{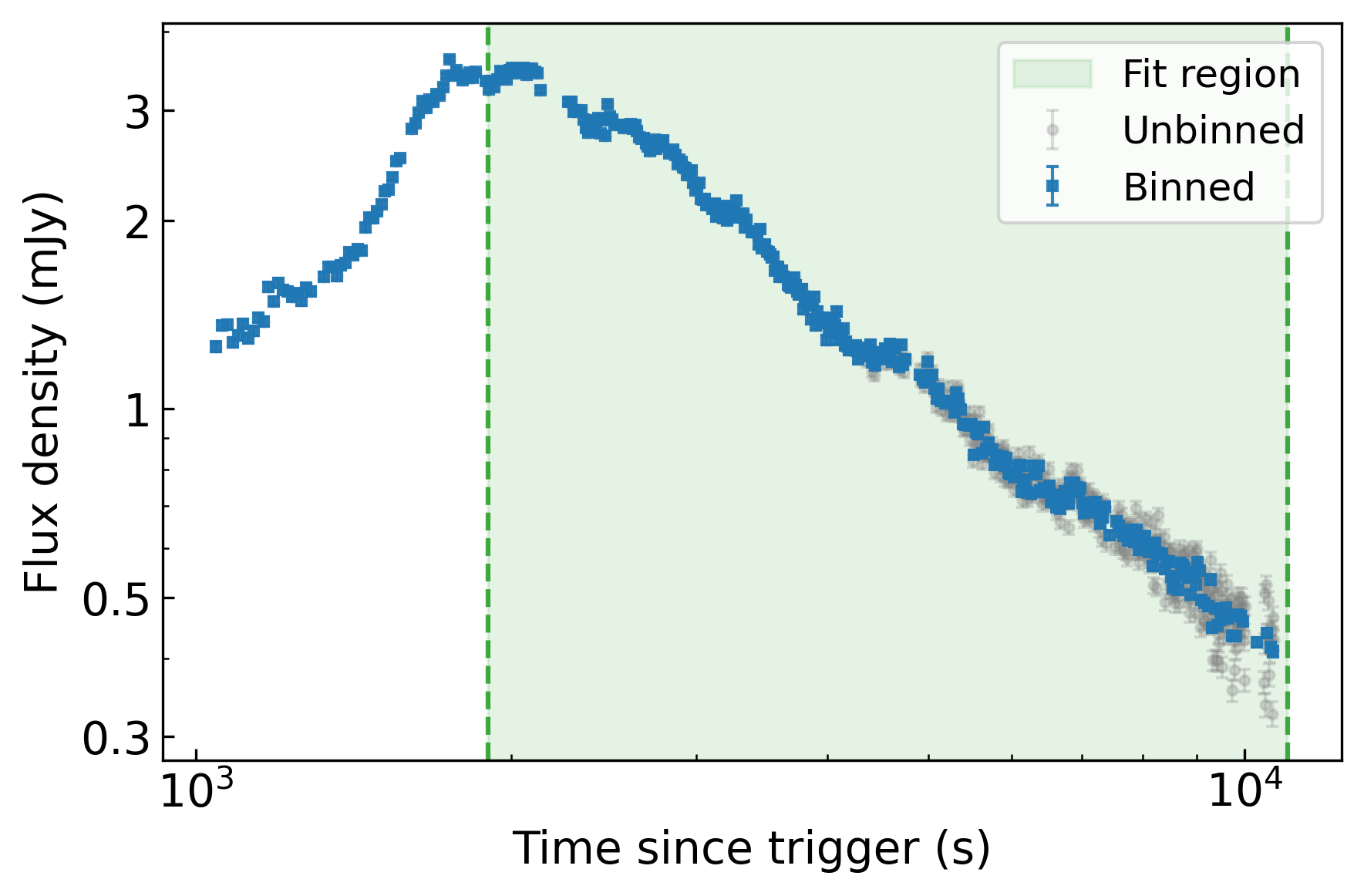}}
    \caption{LAST flux-density light curve of GRB~251013C. Times are transformed to the GRB rest frame and flux is corrected for Galactic extinction. We show the binned (blue squares) and unbinned (gray circles) data. The region used to fit the decay is shown as a green area between two green dashed vertical lines.}
\label{fig:LAST_Wiggles_LC}
\end{figure}

We transformed the observed LAST light curve time scale to the rest frame using the measured redshift of $z = 0.572$ \citep{GCN42227}. We corrected for Galactic extinction assuming $\mathrm{E(B-V)} \simeq 0.0575~$mag, derived from the Planck thermal dust model map \texttt{HFI\_CompMap\_ThermalDustModel\_2048\_R1.20.fits} \citep{Planck_2015_Results}. We adopted the optical extinction curve parameterization of \citet{Extinction} and the Optical Tube Assembly transmission of LAST \citep{LAST_transmission}, assuming a standard extinction parameter of $R_V = 3.1$ and a spectral index of $\beta = 0.75$. Under these assumptions, we obtained an extinction correction of $A_{\mathrm{clear}} = 0.195$~mag. To test the stability of the correction with respect to the assumed $\beta$, we varied $\beta$ over the range $[0,2]$ in steps of $0.01$. The variation resulted in a mean correction of $A_{\mathrm{clear}} = 0.192 \pm 0.006$~mag, indicating that the correction is weakly sensitive to the assumed spectral slope. The extinction-corrected magnitudes were converted to flux density following the AB system, using
\begin{equation}
    F_\nu = 3.631 \times 10^6 \times 10^{-0.4 m_{\mathrm{LAST}}} \ \mathrm{mJy}.
\end{equation}
The complete unbinned LAST photometry used in this work is provided as a machine-readable table.

The light curve was then adaptively rebinned to reduce the scatter at late times. Consecutive measurements were accumulated in time order until the inverse-variance weighted signal-to-noise ratio (S/N) of the bin reached or exceeded a target threshold. The representative bin time was taken as the inverse-variance weighted mean of the contributing measurements. We chose the threshold as the lowest S/N ($=55$) of the pre-peak data points in order to retain the earlier end of the light curve at its finest, unbinned time resolution.

The light curve of GRB~251013C (Fig.~\ref{fig:LAST_Wiggles_LC}) as observed by LAST shows a pronounced rebrightening phase from $\sim1000\,$s to $\sim1900\,$s post-burst, with an increase of $\sim2.36\,$mJy. Given indications of a decaying optical emission prior to the LAST coverage (Fig.~\ref{fig:GRB251013C_lightcurve_with_inset}), we interpret this feature as a rebrightening episode rather than the afterglow onset. 

The rising component of this episode exhibits a break at $\sim1440\,$s, after which the slope steepens significantly. A segmented PL fit yields a rise index of $\alpha = 2.9 \pm 0.2$ following the break, while the post-peak decay is characterized by an index of $\alpha = - 1.32 \pm 0.02$. A smooth broken PL fit places the turnover at $t_{\rm peak} = (1860.0 \pm 20.0)$~s and yields a characteristic width of $\Delta t_\mathrm{FWHM} = (1760.0 \pm 60.0)$~s, corresponding to a fractional timescale $\Delta t/t \sim 1$ for this feature.

The large width and smooth evolution of this rebrightening are consistent with relativistic smoothing on angular timescales (e.g., \citealt{KumarPiran2000, Ioka2005}), as expected for emission from an extended region of the relativistic outflow. In contrast, the decay following the peak exhibits pronounced short-timescale variability superimposed on the overall trend, which we analyze in detail below.

\subsection{Short-timescale deviations}

To characterize the short-timescale variability during the decay phase, we modeled the light curve after the peak ($T-T_0>1900$~s; Fig.~\ref{fig:LAST_Wiggles_LC}) as a PL baseline with a residual component,
\begin{equation}
    F_i = F_{\rm PL}(t_i) + r_i,
\end{equation}
where
\begin{equation}
    F_{\rm PL}(t) = A (t/t_\mathrm{ref})^\alpha .
\end{equation}
Here $r_i$ is the deviation from the PL model at the $i$th observed epoch, in flux density, and $t_\mathrm{ref}=1900~$s. We fitted the PL baseline and residual model jointly. The PL is constrained not to overpredict the observed flux, so the residuals represent excess emission above a monotonic decay. This conservative choice assigns as little of the long-timescale variability to the baseline as possible and therefore yields an upper estimate of the residual correlation scale. Further discussion on the choice of baseline and its effects on the results can be found in App.~\ref{app:baseline}. Baselines spanning the full broad rebrightening were also tested. They recover consistent residual structure during the decay but are poorly constrained during the sparsely sampled pre-kink rise (App.~\ref{app:decayonly}). Before the fitting procedure, the photometric uncertainties were scaled by the correction factor described in App.~\ref{app:errcal}, which accounts for the part of the per-exposure scatter that the pipeline uncertainties do not capture.

To quantify temporal correlations in the residuals without assuming a particular flare shape, we modeled $r_i$ as an autoregressive (AR) process \citep[e.g.,][]{Box2015},
\begin{equation}
    r_i = \sum_{k=1}^{p} \beta_k r_{i-k} + \epsilon_i ,
    \label{eq:ar}
\end{equation}
where $p$ is the AR order, $\beta_k$ are the AR coefficients, and $\epsilon_i$ is the innovation term. In this model, each residual is described in terms of the preceding $p$ residuals, while $\epsilon_i$ represents the part that is not predicted by that preceding sequence. The AR model therefore characterizes the temporal correlation of the residuals without prescribing the shape or number of individual variability features.

After fitting the AR model, we evaluated the implied autocorrelation function (ACF) of the actual fractional time separation, $\Delta t/t$, using the observed sampling. The ACF was obtained exactly from the fitted coefficients through the Yule-Walker relations \citep{Yule1927, Walker1931}.

\begin{figure}
\centerline{\includegraphics[width=0.49\textwidth]{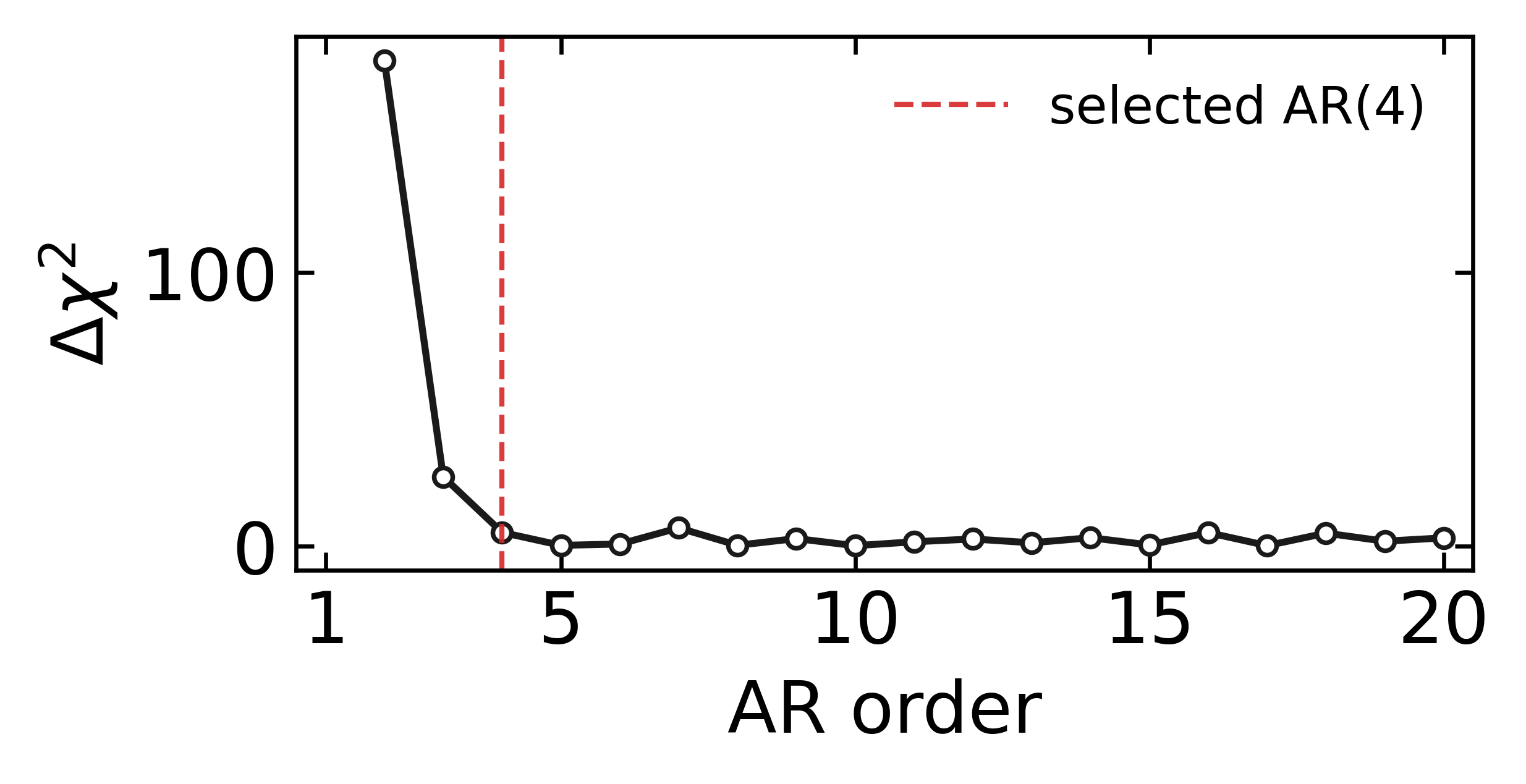}}
    \caption{Autoregression model-order selection. Shown is the improvement $\Delta\chi^2=\chi^2_{\mathrm{AR}(p-1)}-\chi^2_{\mathrm{AR}(p)}$ for adjacent AR orders. The selected order, marked by the vertical red dashed line, is determined using sequential likelihood-ratio tests.}
\label{fig:GRB251013C_AR_order_delta_chi2}
\end{figure}

We fitted AR models over a range of orders up to $p=20$ and selected the order used for the subsequent analysis with a sequential likelihood-ratio test. Each step from AR($p-1$) to AR($p$) adds one coefficient, so we computed
\begin{equation}
\Delta \chi^2_p = \chi^2_{\rm AR(p-1)} - \chi^2_{\rm AR(p)}
\end{equation}
for each adjacent pair of models, evaluated over their common time range. If the additional coefficient is not required, $\Delta \chi^2_p$ is approximately distributed as $\chi^2$ with one degree of freedom ($\chi^2_1$). We therefore defined the likelihood-ratio test p-value as
\begin{equation}
p_{\rm LRT} = P\left(\chi^2_1 \geq \Delta \chi^2_p\right),
\end{equation}
and selected the lowest AR order for which adding the next coefficient no longer gives a significant improvement. With a threshold of $p_{\rm LRT}<0.05$, this procedure selected an AR(4) model (Fig.~\ref{fig:GRB251013C_AR_order_delta_chi2}), with
\begin{equation}
\beta = (0.335, 0.350, 0.193, 0.100).
\end{equation}
The largest coefficients occur at the first few lags, indicating short-range temporal correlations in the deviations from the PL decay. The fractional residuals (Fig.~\ref{fig:GRB251013C_LC_PLandFREDs_fit}) about the PL model have an rms of $\sigma_r = 0.0774,$ whereas the AR innovations have $\sigma_\epsilon = 0.0359$. Thus the innovation scatter is $0.46$ of the residual rms, indicating that much of the observed variability is temporally correlated rather than independent measurement scatter.

We applied two tests to establish that the deviations are correlated in time. If the deviations were independent, the coefficients of Eq.~(\ref{eq:ar}) would all be zero, whereas fitting the four coefficients lowers $\chi^2$ by $2236$. Under the null hypothesis $\Delta\chi^2$ follows approximately a $\chi^2$ distribution with four degrees of freedom, giving a formal equivalent Gaussian significance of $\simeq47\sigma$. The precise value carries little meaning, since it requires the asymptotic $\chi^2$ form to hold at a $p$-value of order $10^{-480}$, but the null hypothesis is nonetheless overwhelmingly rejected. The second test uses no uncertainties at all. Residuals of a common sign about the median occur in stretches of $4.7$ time samples on average, against $2.0$ expected for an independent sequence, so that independence is rejected at $10.7\sigma$. In both tests we first removed a baseline-proportional offset, so that they measure the fluctuation about the baseline rather than the overall excess above it.

\begin{figure}
\centerline{\includegraphics[width=0.49\textwidth]{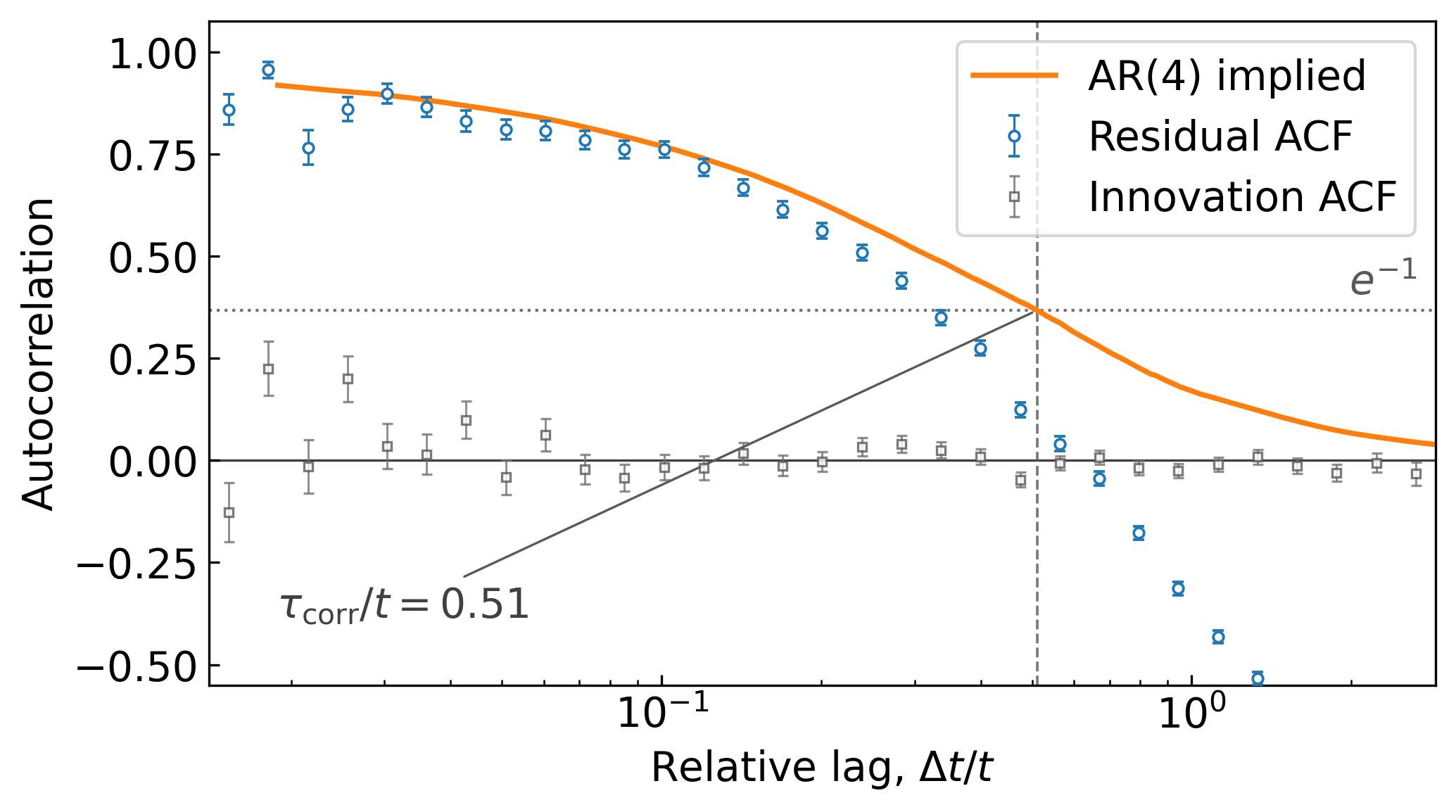}}
    \caption{Autocorrelation function (ACF) of the PL residuals. Blue points show the empirical residual ACF binned by relative lag, \(\Delta t/t\), while the orange curve shows the ACF implied by the best-fit AR(4) residual model. Gray squares show the ACF of the AR innovations, which are approximately decorrelated after applying the AR model. The horizontal dotted line marks \(e^{-1}\), and the vertical dashed line indicates the AR-implied decorrelation scale, \(t_{\rm corr}/t \simeq 0.51\).}
\label{fig:GRB251013C_acf_paper}
\end{figure}

We estimated the characteristic timescale of these correlated deviations from the autocorrelation function implied by the AR model using logarithmic $\Delta t/t$ lags (Fig.~\ref{fig:GRB251013C_acf_paper}). We defined the decorrelation scale as the first lag at which the absolute autocorrelation falls below $1/e$. For the selected AR(4) model, this occurs at
\begin{equation} \label{eq:decorrelationtime}
    \frac{t_{\rm corr}}{t} \simeq 0.51.
\end{equation}

The decorrelation scale characterizes the residual variability relative to the adopted baseline. The result shows that the PL deviations decorrelate on a fractional timescale of order a few tenths, shorter than the broad rebrightening scale of $\Delta t/t\sim1$. Robustness tests for the AR order, residual ordering, time-sampling cadence, and noise-only realizations are presented in App.~\ref{app:ar_robustness}.

\subsection{Variability decomposition}
\label{sec:FREDdecomposition}

\begin{figure*}
\centering
\begin{minipage}[t]{0.70\textwidth}
    \vspace{0pt}
    \centering
    \includegraphics[width=\textwidth]{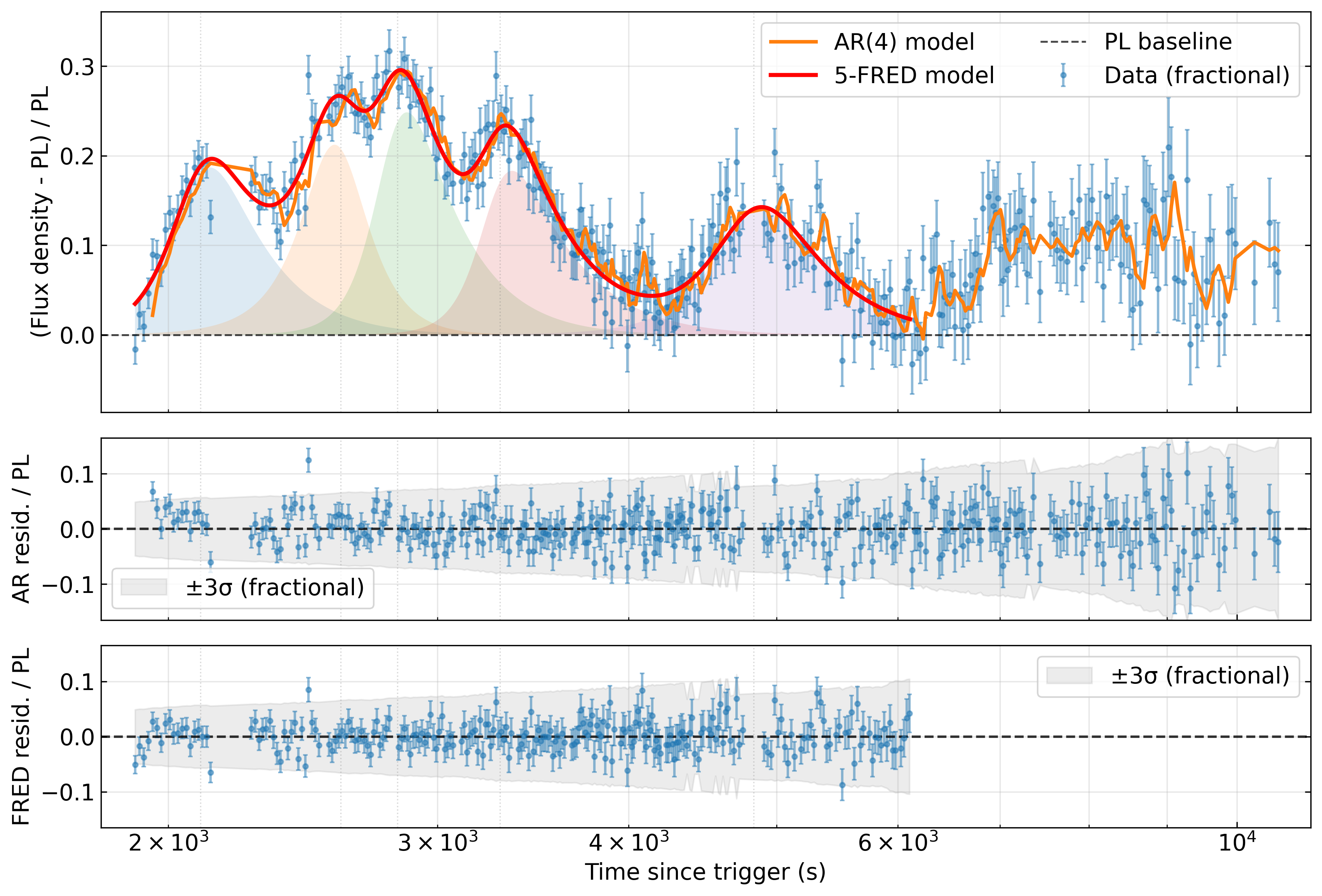}
\end{minipage}
\hfill
\begin{minipage}[t]{0.26\textwidth}
    \vspace{0pt}
    \caption{Fractional post-peak light curve of GRB~251013C from LAST observations, relative to the baseline power-law (PL) decay. Shown are the data with $1\sigma$ uncertainties (blue data points) after applying the field-star correction (App.~\ref{app:errcal}), the best-fit AR model (orange jagged solid line), the best-fit composite FRED model (smooth red line), and the individual FRED contributions (filled areas). The bottom panels show fractional model residuals relative to the PL baseline for both AR and FRED models, with the $\pm3\sigma$ standard deviation bands in gray.}
    \label{fig:GRB251013C_LC_PLandFREDs_fit}
\end{minipage}
\end{figure*}

The AR analysis measures the correlation scale of the residuals without identifying individual structures. As a complementary description, we fitted the post-peak residual emission with a sum of phenomenological fast-rise exponential-decay (FRED) components. We fitted the fractional residuals $r_i/F_\mathrm{PL}(t_i)$ during the high-S/N part of the decay phase, $1900~{\rm s}<T-T_0<6100~{\rm s}$, where the residual light curve contains several well-resolved local maxima.

The model consists of a sum of superimposed FRED components,
\begin{equation}
R(t) = \sum_{i=1}^{N} 
A_i \frac{\exp\!\left[-\frac{t - t_{p,i}}{\tau_{d,i}}\right]}
{1 + \exp\!\left[-\frac{t - t_{p,i}}{\tau_{r,i}}\right]},
\end{equation}
where $A_i$, $t_{p,i}$, $\tau_{r,i}$, and $\tau_{d,i}$ denote the amplitude, peak time, rise timescale, and decay timescale of the $i$th component, respectively. The peak times $t_p$ are given in the rest-frame post-trigger time. We chose five components to describe the clearly resolved peaks and fitted them simultaneously (Fig.~\ref{fig:GRB251013C_LC_PLandFREDs_fit}), allowing the peak times to vary within $\pm1\%$ of their selected values. Because neighboring features overlap, the fitted FREDs are not assumed to correspond one-to-one to distinct physical emission episodes. The purpose is instead to characterize the typical width of the resolved substructure.

\begin{table*}
\centering
\caption{Fitted FRED pulse parameters and derived episode properties.}
\label{tab:FREDparams}
\begin{tabular}{lcccc|cccc}
\hline
 & \multicolumn{4}{c|}{fitted parameters} & \multicolumn{4}{c}{derived properties} \\
Peak & $t_p$ (s) & $A$ & $\tau_r$ (s) & $\tau_d$ (s) & $\Delta t_{\rm FRED}$ (s) & $\Delta t_{\rm FRED}/t_p$ & amplitude (mJy) & significance ($\sigma$) \\
\hline
1 & $2079\pm46$  & $0.34\pm0.05$ & $55\pm8$   & $191\pm132$ & $663\pm187$  & $0.32\pm0.09$ & $0.53\pm0.08$ & $12.8$ \\
2 & $2568\pm119$ & $0.42\pm0.17$ & $58\pm29$  & $118\pm220$ & $588\pm314$  & $0.23\pm0.12$ & $0.48\pm0.20$ & $5.4$ \\
3 & $2797\pm66$  & $0.44\pm0.44$ & $62\pm46$  & $249\pm266$ & $812\pm381$  & $0.29\pm0.14$ & $0.49\pm0.28$ & $4.7$ \\
4 & $3263\pm27$  & $0.31\pm0.13$ & $71\pm25$  & $337\pm57$  & $1038\pm88$  & $0.32\pm0.03$ & $0.30\pm0.10$ & $ 10.6$ \\
5 & $4782\pm85$  & $0.26\pm0.02$ & $163\pm21$ & $489\pm91$  & $1828\pm132$ & $0.38\pm0.03$ & $0.14\pm0.01$ & $ 15.4$ \\
\hline
\end{tabular}
\tablefoot{Peak times are in the rest-frame post-trigger time. The amplitude is the peak excess of the component over the PL baseline. Because that baseline is constrained to the lower envelope, the amplitudes carry a common offset of about $0.03$ in fractional flux. Significances follow from removing the component and refitting the remainder, evaluated against a simulated null distribution. For the overlapping components 2--4, the data constrain their combined shape better than the individual amplitudes, so a large uncertainty on one amplitude reflects how the flux divides between neighbors rather than whether the component is required.}
\end{table*}

For each component, we derived characteristic timescales from the fitted parameters (Tab.~\ref{tab:FREDparams}). We consider the rise time $t_\mathrm{rise}$ (10--90\%) and decay time $t_\mathrm{decay}$ (90--10\%) derived from the fitted profiles as empirical measures of the temporal extent of each variability feature. We define the total FRED width as
\begin{equation}
    \Delta t_{\rm FRED} = t_{\rm rise} + t_{\rm decay},
    \label{eq:fredwidth}
\end{equation}
which avoids dependence on the less sharply defined far tails of the profiles.

Table~\ref{tab:FREDparams} also lists each component's width, its ratio to the peak time, its peak amplitude, and a significance. The significance was obtained by removing one component and refitting the remaining four. We calibrated against simulations rather than a tabulated $\chi^2$, since a component whose amplitude vanishes leaves its width and position undefined. Fitting the $\Delta\chi^2$ distribution of 2000 null realizations with a $\chi^2$ law gives 3.4 effective degrees of freedom. All five components are required, at $12.8$, $5.4$, $4.7$, $10.6$ and $15.4\sigma$.

\begin{figure}
\centerline{\includegraphics[width=0.49\textwidth]{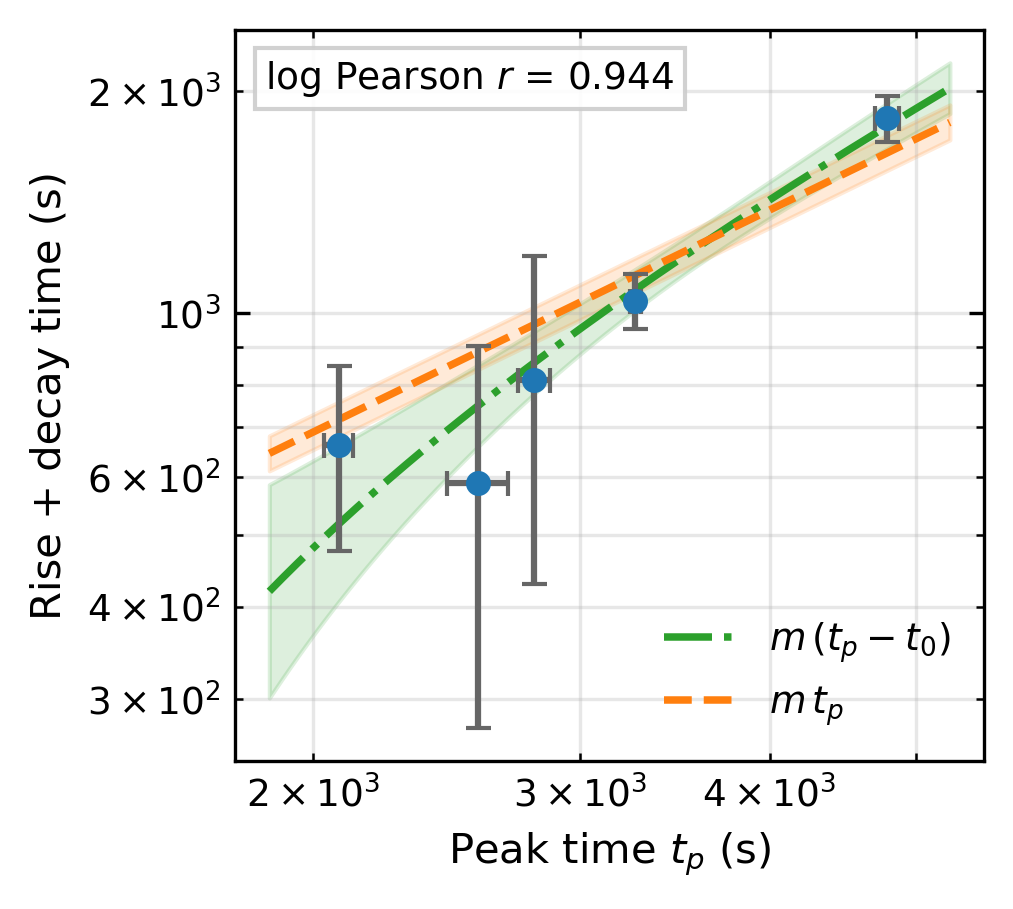}}
    \caption{Evolution of the characteristic FRED widths as a function of peak time. The FRED widths (data points) were fitted with a scale-free proportional model $mt_p$, which we adopt, and with a linear-offset model $m(t_p-t_0)$, shown for comparison. The offset model improves on the proportional one at only $1.3\sigma$ and is not adopted.}
\label{fig:GRB251013C_timescales}
\end{figure}

We tested whether the fitted component widths increase with peak time (Fig.~\ref{fig:GRB251013C_timescales}) by comparing a constant-width model, $\Delta t_{\rm FRED} = a$, which corresponds to variability drawn from a fixed absolute timescale, with a scale-free proportional relation, $\Delta t_{\rm FRED} = m\,t_p$, which represents a constant fractional variability timescale measured from the GRB trigger. We used a Gaussian likelihood, taking the effective uncertainty on each width to include the contribution of the peak-time uncertainty through the model slope. The proportional fit gives
\begin{equation}
m = 0.344 \pm 0.019,
\end{equation}
with $\chi^2/\mathrm{d.o.f.}=3.873/4$. Because the two models are not nested, we compared them with the Akaike information criterion (AIC; \citealt{HurvichTsai1993}) in its finite-sample form for $n=5$. The proportional description is strongly preferred over the constant-width model, with $\Delta{\rm AIC}\simeq35$. We also tested a linear relation  $\Delta t_{\rm FRED} = m(t_p-t_0)$ in which the effective time origin $t_0$ is allowed to vary, which gives $m = 0.47 \pm 0.08$ and $t_0 = (0.97 \pm 0.44)\times10^{3}$~s with $\chi^2/\mathrm{d.o.f.}=1.110/3$. It improves on the proportional model at only $1.3\sigma$ and the fitted offset differs from the trigger time by $2.2\sigma$, so we do not adopt it.

While the broad rebrightening has a duration comparable to its observer time, the resolved FRED-like residual structures have shorter characteristic timescales, with $\Delta t_{\rm FRED}/t_p < 1$. Together with the observed width growth, this supports a two-scale empirical description of the LAST light curve.

\subsection{Optical-to-X-ray spectral energy distribution}
\label{sec:opticalxray}

\begin{figure}
\centering
\includegraphics[width=\columnwidth]{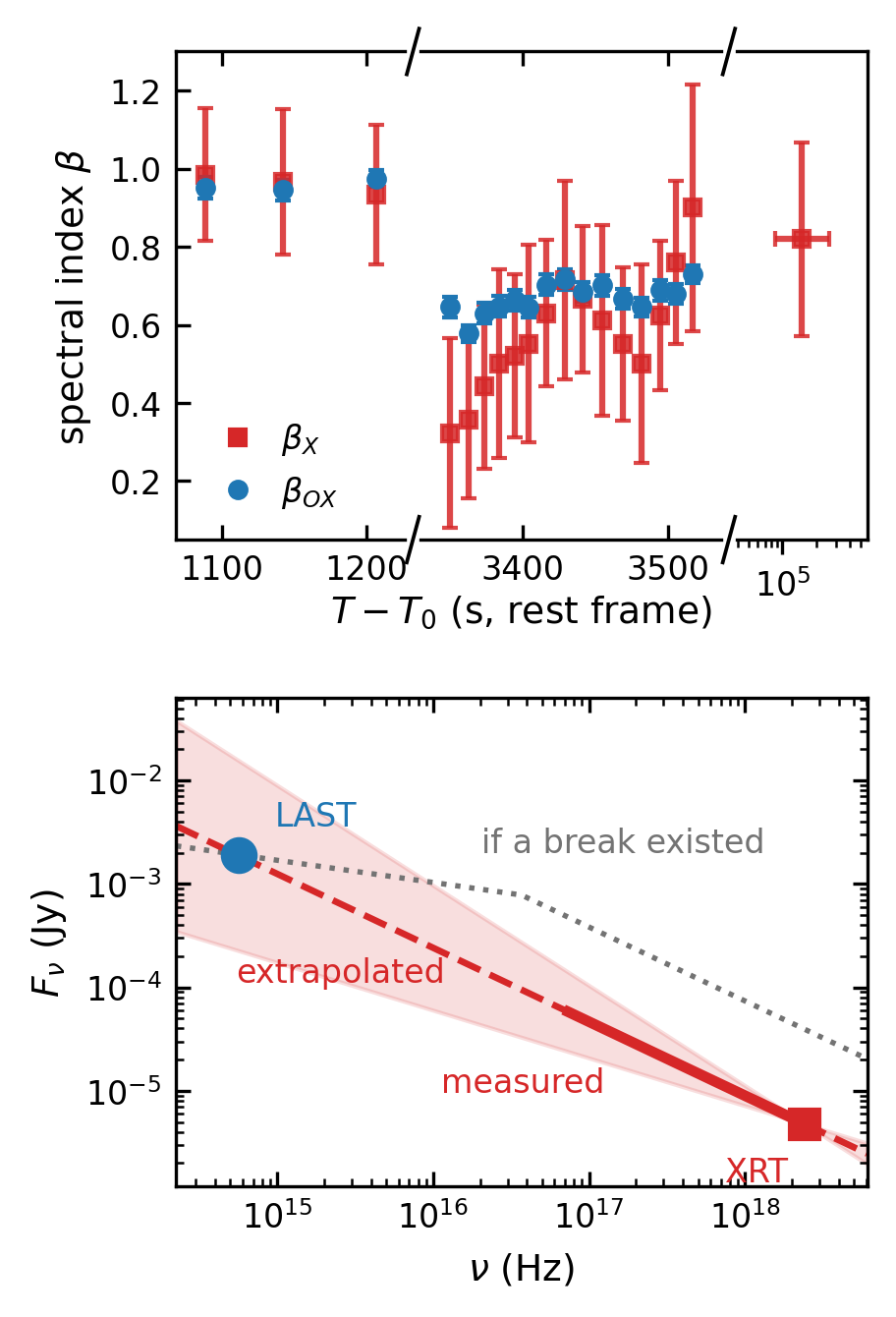}
\caption{Optical-to-X-ray spectral energy distribution. The top panel shows the in-band X-ray index $\beta_X$ (red squares) and optical-to-X-ray index $\beta_{OX}$ (blue circles) for overlapping XRT and LAST epochs. The rightmost point is the late X-ray epoch, where a single spectral fit covers two flux bins. The bottom panel shows the spectrum at $T-T_0 = 3429$~s. The line and its $\pm1\sigma$ band are set by the X-ray alone, and the optical point is the prediction being tested. Also shown is an illustrative spectrum with a cooling break (dotted line).}
\label{fig:optical_Xray_SED}
\end{figure}

The X-ray afterglow was observed with the X-Ray Telescope (XRT; \citealt{Burrows2005}) on board the Neil Gehrels Swift Observatory (\textit{Swift}; \citealt{Gehrels2004}). We used the time-resolved XRT products \citep{Evans2009}, whose light curve is the unabsorbed flux density at $10$~keV. Two intervals fall inside the LAST coverage, $1089$ to $1207$~s and $3349$ to $3517$~s in the rest frame, with 3 and 15 epochs. The first precedes the broad rebrightening, which peaks at $1860$~s, and the second falls on the decay of the 4th FRED episode.

A likelihood-ratio test on the photon indices, comparing a single mean across all epochs against a separate mean for the flare, gives $\Gamma_{X,\rm out} = 1.89 \pm 0.04$ and $\Gamma_{X,\rm flr} = 1.57 \pm 0.06$, with a significance of $4.5\sigma$, so the XRT spectrum hardens during the flare. To test whether the optical lies on that spectrum we compared the in-band index $\beta_X = \Gamma_X - 1$ with the slope joining the dereddened LAST flux density to the X-ray point,
\begin{equation}
    \beta_{OX} = -\,\frac{\log\left(F_{\nu,O}/F_{\nu,X}\right)}{\log\left(\nu_O/\nu_X\right)},
\end{equation}
where $F_{\nu,O}$ and $F_{\nu,X}$ are the dereddened LAST and unabsorbed XRT flux densities at the effective frequencies $\nu_O$ and $\nu_X$ of the two bands. Here $\nu_O = 5.62\times10^{14}$~Hz is the effective frequency of the unfiltered LAST band, weighted by the system throughput under the same spectral assumption used for the extinction correction, and $\nu_X = 2.42\times10^{18}$~Hz corresponds to $10$~keV. Both indices agree at both epochs (Fig.~\ref{fig:optical_Xray_SED}), differing by $-0.005\pm0.104$ before the flare and $+0.097\pm0.057$ during it.

A cooling break between the bands would shift $\beta_{OX}$ below $\beta_X$ by $0.5u$, where $u$ is its fractional position across the $3.63$ decades separating them. The measured offset confines any such break to within $0.17$ decades of the LAST band at $2\sigma$, that is to the lowest $5\%$ of the range, and to $0.61$ decades at $3\sigma$. A break below the optical band produces no offset and is not constrained by this test.

Within the X-ray band $\beta_X$ falls by $0.392\pm0.118$, and from optical-to-X-ray $\beta_{OX}$ falls by $0.292\pm0.018$, agreeing at $0.8\sigma$, showing a consistent hardening across the broadband spectrum. Both indices are constant through the first interval. The spectrum hardens somewhere in the gap between the intervals, and within the second it softens again, $\beta_{OX}$ rising from $0.65$ to $0.73$ at $4.1\sigma$ as the flare decays.

A PL that changes slope while its optical end stays nearly fixed must change flux at its X-ray end. Between the intervals the optical rises by a factor of $1.34$ while $\beta_{OX}$ hardens by $0.292$. Together these require the X-ray to rise by a factor of $15.5$, whereas the observed rise is $15.7$. The X-ray brightening is therefore consistent with a spectral change.

\section{Interpretation}
\label{sec:interpretation}

Several mechanisms could in principle contribute to the observed morphology. Late internal dissipation could produce rapid variability and spectral evolution, but would require an engine episode with a duration comparable to the observer time, while circumburst density structure is less well suited to producing the observed sharp substructure and broadband hardening \citep{Nakar2003, Ioka2005, NakarGranot2007, Zhang2006, Lazzati2007}. Passage of a synchrotron spectral break alone also cannot account for the large-amplitude rebrightening and its superposed fluctuations. We therefore adopt a refreshed external-shock interpretation as the simplest working scenario and examine the physical conditions required to reproduce the observations.

\subsection{A refreshed external-shock interpretation}
\label{sec:refreshed_shocks}

A broad optical rebrightening can be produced by delayed energy injection into the external shock. In the refreshed-shock scenario, slower shells catch up with the decelerating blast wave at later times, increasing the blast-wave energy and producing a flattening or rebrightening of the afterglow \citep{ReesMeszaros1998, SariMeszaros2000, KumarPiran2000, ZhangMeszaros2002, Granot2003, Moss2023}. This does not necessarily require central-engine activity lasting until the time of the bump, since the delayed collision may instead reflect an initial distribution of ejecta Lorentz factors \citep{ReesMeszaros1998, SariMeszaros2000, ZhangMeszaros2002}.

We interpret the main optical bump as a substantial, structured refreshed interaction rather than continuous energy injection from a smooth ejecta distribution. Its characteristic duration, $\Delta t/t\sim1$, is consistent with the angular and dynamical timescales of an external shock \citep{Nakar2003, Ioka2005}. The additional broad episodes before and after the LAST observations (Fig.~\ref{fig:GRB251013C_lightcurve_with_inset}) may likewise trace successive collisions with slower ejecta.

The broadband spectrum provides a constraint on the emitting electron population. During the sampled flare interval, the optical flux is consistent with an extrapolation of the X-ray PL, while the spectrum hardens relative to the pre-flare epoch. We therefore consider a newly dominant synchrotron population associated with the refreshed interaction, rather than assigning the optical bump and X-ray flare to unrelated emission mechanisms. In our fiducial interpretation, the refreshed forward shock dominates the broad emission, while the pre-existing afterglow contributes an underlying component whose temporal evolution is not independently determined.

\subsection{Broadband spectrum and circumburst environment}
\label{sec:refreshed_shocks_spectrum}

The hard flare spectrum alone does not uniquely determine the cooling regime. 
In the fast-cooling ordering $\nu_c<\nu_{\rm opt}<\nu_X<\nu_m$, the expected spectral index is $\beta=1/2$, close to the observed value. However, this regime predicts a temporal decay $\alpha=-1/4$ for both ISM- and wind-like environments, far shallower than the observed post-peak decay, $\alpha=-1.32\pm0.02$. If instead both bands lie above $\nu_m$ in the fast-cooling regime, $\beta=p/2$ implies $p\simeq1.14$ and $\alpha=-(3p-2)/4\simeq-0.36$, which is likewise inconsistent with the observed decay. We therefore adopt the slow-cooling ordering
\begin{equation}
\nu_m < \nu_{\rm opt} < \nu_X < \nu_c
\end{equation}
for the dominant flare emission. In this regime, $\beta=(p-1)/2$, and the hard X-ray photon index $\Gamma_{X,\rm flr}=1.57\pm0.06$ implies
\begin{equation}
p_{\rm flr}=2\Gamma_{X,\rm flr}-1
=2.14\pm0.12.
\end{equation}
The corresponding standard forward-shock closure relations are
\begin{equation}
\alpha_{\rm ISM}=-\frac{3(p-1)}{4},
\qquad
\alpha_{\rm wind}=-\frac{3p-1}{4}.
\end{equation}
For $p=2.14$, these predict $\alpha_{\rm ISM}\simeq-0.86$ and $\alpha_{\rm wind}\simeq-1.36$. The latter agrees closely with the observed post-peak decay, whereas the ISM prediction is substantially shallower. We therefore favor a wind-like circumburst medium for the dominant forward-shock component, subject to the assumptions of an adiabatic blast wave, approximately constant microphysics, and negligible continuing energy injection during the fitted decay.

The adopted spectral ordering requires the cooling frequency to remain above the X-ray band. In a wind medium,
\begin{equation}
\nu_c\propto E^{1/2}A_*^{-2}\epsilon_B^{-3/2}
t^{1/2}(1+Y)^{-2},
\end{equation}
where $E$ is the isotropic-equivalent kinetic energy of the blast wave, $A_*$ is the wind-density normalization, $\epsilon_B$ is the fraction of the post-shock energy carried by the magnetic field, $t$ is the rest-frame time since the burst, and $Y$ is the Compton parameter \citep{ChevalierLi2000, GranotSari2002, Gao2013}. A sufficiently low wind density or magnetic energy fraction can therefore accommodate a high cooling frequency, although the absolute values cannot be determined from the present optical and X-ray data alone.

The spectral hardening places an additional requirement on the refreshed interaction. If both epochs occupy the same uncooled segment, the out-of-flare photon index $\Gamma_{X,\rm out}=1.89\pm0.04$ corresponds to
\begin{equation}
p_{\rm out}=2.78\pm0.08,
\end{equation}
compared with $p_{\rm flr}=2.14\pm0.12$. Energy injection with a fixed electron index changes the flux normalization and characteristic frequencies, but does not by itself produce this change in spectral slope. We therefore require the newly dominant emission to have a harder electron distribution than the pre-flare emission. A structured collision can modify the shock strength, magnetization, and acceleration conditions, and may also generate a reverse shock that accelerates a distinct electron population. The data do not determine which shock produces the harder electrons, and we do not attempt to predict $p$ from the collision dynamics. We therefore treat the required change as an observational constraint on the acceleration process.

The nearly unbroken optical-to-X-ray spectrum is consistent with the harder population dominating both bands during the sampled flare interval. A weaker underlying afterglow component may still be present. However, the XRT coverage misses the main optical peak and most of the fitted fast optical substructures, so the available data do not establish whether those features are accompanied by corresponding X-ray variability.

\subsection{Energy injection and rapid substructure}
\label{sec:refreshed_shocks_energy}

The steep rise of the bump provides an estimate of the strength of the
refreshed interaction. For a wind forward shock with fixed microphysics and
$\nu_m<\nu<\nu_c$, the standard synchrotron scalings give
\citep{ChevalierLi2000, GranotSari2002, Gao2013}
\begin{equation}
    F_\nu\propto E^{(p+1)/4}t^{(1-3p)/4}.
\end{equation}
Writing the blast-wave energy as $E\propto t^{q}$, where $q$ is the
energy-injection index, then gives
\begin{equation}
    \alpha_{\rm rise}
    =\frac{p+1}{4}q-\frac{3p-1}{4}.
\end{equation}
With $p=2.14 \pm 0.12$ and the measured post-kink rise $\alpha_{\rm rise}=2.9\pm0.2$,
this yields $q=5.4\pm0.3$. Between approximately 1440 and 1860 s, the corresponding
illustrative energy ratio is
\begin{equation}
\frac{E_{\rm peak}}{E_{\rm kink}}
\simeq\left(\frac{1860}{1440}\right)^{5.4}
\simeq4.
\end{equation}
This indicates that a substantial energy increase would be required if the rise were attributed entirely to a forward shock with fixed microphysics. However, the estimate is not a measurement of the incoming shell energy. The observed hardening implies a changing electron population, which can modify the optical flux independently of the energy increase, and the scaling is only approximate during a strong structured collision. We therefore use this calculation as an illustrative measure of the strength of the rise, without assigning a precise energy or Lorentz-factor contrast to the incoming shell.

The resolved faster variability provides a further constraint on the structure of this interaction. The fitted components have $\Delta t/t\simeq0.23$--$0.38$, substantially shorter than the broad rebrightening. Such structure is not expected from a smooth, spherically symmetric refreshed forward shock, whose emission is averaged over the visible equal-arrival-time surface \citep{Nakar2003, Ioka2005}. We therefore require the refreshed interaction to be non-smooth. Angular inhomogeneity in the incoming ejecta could modulate the forward-shock emission over only part of the visible surface, while radial structure and a reverse-shock contribution may produce additional variability associated with the arriving material \citep{KumarPiran2000, Granot2003, NakarOren2004, Ioka2005, Moss2023, GaoMeszaros2015}. The present data do not uniquely distinguish these possibilities, so we do not assign the individual phenomenological components to separate physical shocks.

If the shorter variability is dominated by angular light-travel-time effects, a patch of angular scale $\theta_{\rm patch}$ produces, to order unity,
\begin{equation}
    \frac{\Delta t}{t}\sim
    \left(\Gamma\theta_{\rm patch}\right)^2 .
\end{equation}
The measured $\Delta t/t\simeq0.23$--$0.38$ therefore corresponds to
$\Gamma\theta_{\rm patch}\sim0.48$--$0.62$, implying angular structure on a scale of roughly half the visible $1/\Gamma$ region. This estimate is only indicative, since radial structure in the ejecta and the location of the emitting patch within the visible surface can also contribute to the observed duration.

\subsection{Bulk Lorentz factor ordering}
\label{sec:gamma_stratification}

If the ejected matter is radially stratified in Lorentz factor, progressively slower shells can catch up with the decelerating blast wave at later times, producing successive refreshed shocks. In this interpretation, the three broad rebrightening episodes in the full optical light curve (Fig.~\ref{fig:GRB251013C_lightcurve_with_inset}) may trace such collisions. At the time of each collision, the Lorentz factor of the catching-up shell $\Gamma_{\mathrm{sh},i}$ is approximately that of the blast wave $\Gamma_{\rm bw}$ immediately before the interaction,
\begin{equation}
    \Gamma_{{\rm sh},i}\sim\Gamma_{\rm bw}(t_i).
\end{equation}

\begin{figure}
    \centerline{\includegraphics[width=0.49\textwidth]{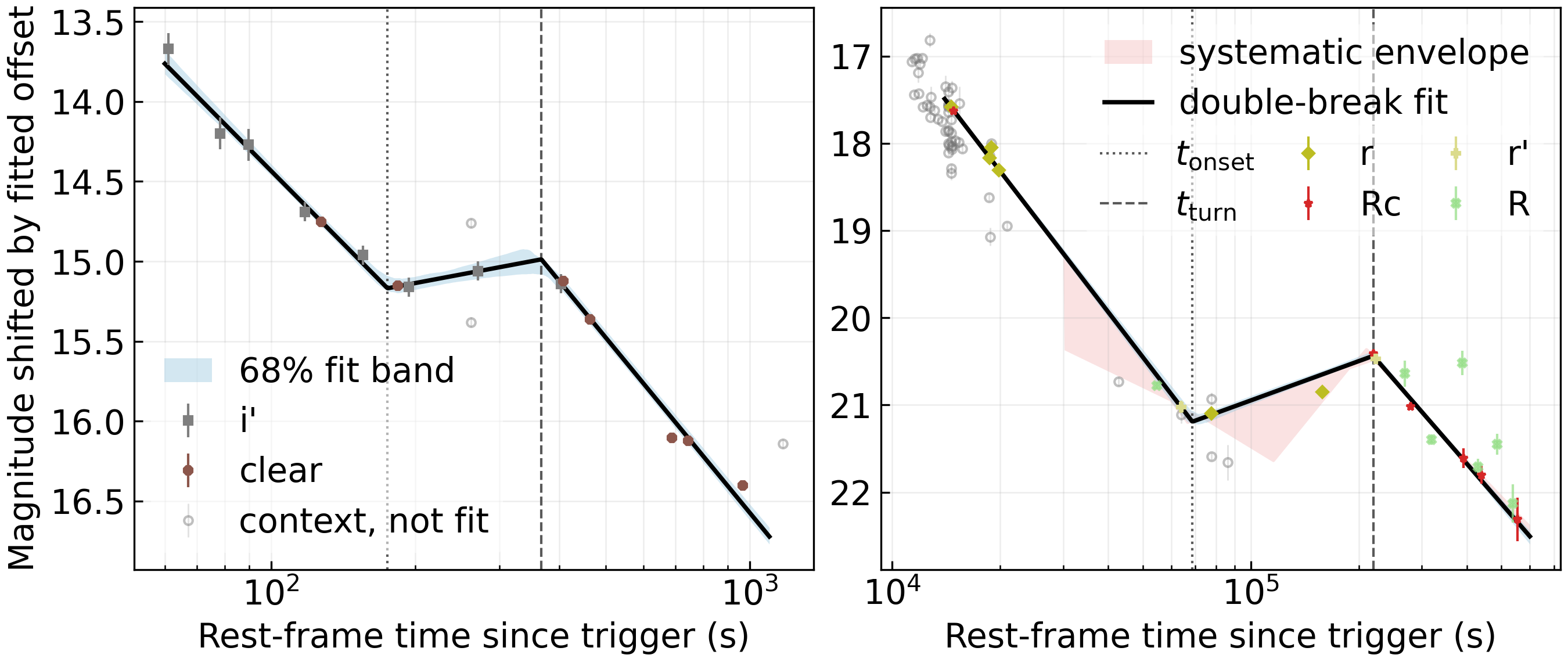}}
    \caption{Continuous double-break fits to the optical rebrightenings before and after the LAST observations. Colored points are included in the fits, while gray points show measurements that were not fitted. The onset $t_{\rm onset}$ and turnover $t_{\rm turn}$ times are marked with dotted and dashed vertical lines, respectively. Blue bands show the central $68\%$ residual-bootstrap intervals. For the late bump, the pink band shows the point-wise envelope of the systematic-variant model curves.}
    \label{fig:other_bumps_estimate}
\end{figure}

To estimate the characteristic times of the optical bumps outside the LAST observing window, we fitted the pre-LAST and post-LAST rebrightenings in rest-frame time. Because the available coverage is heterogeneous, we do not attempt to decompose the light curves into an underlying afterglow and an additive bump. Instead, we fitted a continuous doubly broken PL in magnitude space, allowing independent magnitude offsets between filters (Fig.~\ref{fig:other_bumps_estimate}). The two breaks mark the onset and turnover of each rebrightening, and only the fitted turnover times are used in the Lorentz-factor estimate.

The nominal early-bump fit uses the clear and $i'$ measurements between $60$ and $1100$~s, while the late-bump fit uses the $r$, $r'$, and $R_{\rm C}$ measurements between $1.4\times10^4$ and $6\times10^5$~s. Statistical uncertainties were estimated from 500 residual-bootstrap realizations \citep{EfronTibshirani1993}. For the more sparsely sampled late bump, we additionally estimated a systematic uncertainty from reasonable changes to the fitted data selection. These changes comprised moving the start of the fit from $1.4\times10^4$ to $3\times10^4$~s, adding the $R$-band measurements, and applying both changes together.

The fitted rest-frame turnover times are
\begin{align}
t_{{\rm turn},1}
&=367\pm25_{\rm stat}~{\rm s},\\
t_{{\rm turn},2}
&=1860\pm20~{\rm s},\\
t_{{\rm turn},3}
&=\left(
21.95\pm0.47_{\rm stat}
\,{}^{+0.00}_{-1.01}{}_{\rm syst}
\right)\times10^4~{\rm s}.
\end{align}
Assuming approximately symmetric rise and decay durations, the normalized full-duration estimates are
\begin{align}
    \frac{\Delta t_1}{t_{{\rm turn},1}}
    &= 1.05 \pm 0.09_{\rm stat},\\
    \frac{\Delta t_3}{t_{{\rm turn},3}}
    &= 1.37 \pm 0.03_{\rm stat}
    \,{}^{+0.02}_{-0.50}{}_{\rm syst}.
\end{align}
We use these only as an order-of-magnitude check that the modulations are broad.

The spectral and temporal closure relations (Sect.~\ref{sec:refreshed_shocks_spectrum}) favor a wind-like circumburst medium, which we adopt as the fiducial environment. For
$\rho=Ar^{-2}$, with $A=5\times10^{11}A_*~{\rm g~cm^{-1}}$, the blast-wave Lorentz factor evolves as
\begin{equation}
    \Gamma_{\rm bw}(t_{\rm rest}) \simeq
    6.0
    \left(\frac{E_{\rm k,iso,53}}{A_*}\right)^{1/4}
    \left(\frac{t_{\rm rest}}{1~{\rm day}}\right)^{-1/4}.
    \label{eq:wind_Gamma}
\end{equation}

The measured prompt isotropic energy is $E_{\gamma,\rm iso}=5.6\times10^{51}$~erg \citep{GCN42695}. For a fiducial efficiency $\eta_\gamma=0.2$, this corresponds to a reference kinetic energy $E_{\rm k,iso,0}\simeq2.2\times10^{52}$~erg. Since refreshed shocks can increase the blast-wave energy between episodes, we denote the kinetic energy immediately before collision $i$ by $E_{{\rm k},i}$. Equation~(\ref{eq:wind_Gamma}) then gives
\begin{align}
    \Gamma_1 &\sim
    16\,A_*^{-1/4}
    \left(\frac{E_{{\rm k},1}}{2.2\times10^{52}~{\rm erg}}\right)^{1/4},\\
    \Gamma_2 &\sim
    11\,A_*^{-1/4}
    \left(\frac{E_{{\rm k},2}}{2.2\times10^{52}~{\rm erg}}\right)^{1/4},\\
    \Gamma_3 &\sim
    3.3\,A_*^{-1/4}
    \left(\frac{E_{{\rm k},3}}{2.2\times10^{52}~{\rm erg}}\right)^{1/4}.
\end{align}

Because $\Gamma_{\rm bw}\propto(E_{\rm k}/t)^{1/4}$, the turnover times alone do not uniquely determine the ordering if the blast-wave energy changes substantially. The conditions for progressively decreasing Lorentz factor are
\begin{equation}
    \frac{E_{{\rm k},2}}{E_{{\rm k},1}}<5.1,
    \qquad
    \frac{E_{{\rm k},3}}{E_{{\rm k},2}}<118.
\end{equation}
The second condition is therefore very weak. The first is more restrictive, and the energy added by the first collision is not constrained by the present data. The characteristic times are consistent with progressively slower catching-up ejecta provided that collision does not raise the blast-wave energy by more than a factor of $\simeq 5.1$, so the numerical Lorentz factors and their ordering remain conditional on the energy evolution.

\section{Discussion}
\label{sec:discussion}

The primary observational result of this work is that the early optical afterglow of GRB~251013C exhibits variability on at least two temporal scales. The light curve shows broad rebrightening episodes with $\Delta t/t\sim1$ modulated by shorter-timescale substructure with $\Delta t/t<1$. The AR and FRED analyses provide complementary empirical characterizations of this residual variability, but do not require assigning a unique physical origin to each feature. Contemporaneous optical and X-ray observations additionally show significant broadband spectral hardening during the main rebrightening.

This separation of timescales is important for interpreting the emission. The broad rebrightening is consistent with variability from an extended external-shock region, where equal-arrival-time effects and angular smoothing naturally produce features with durations on the order of the observer time. We therefore interpret the main rebrightening as a substantial refreshed interaction. The spectral and temporal behavior favors a wind-like medium in the slow-cooling regime $\nu_m<\nu_{\rm opt}<\nu_X<\nu_c$. The flare spectrum then implies $p\simeq2.14$, for which the standard wind closure relation predicts a post-peak decay consistent with the observed $\alpha=-1.32\pm0.02$.

The spectral hardening further requires the refreshed interaction to produce a newly dominant, harder electron population rather than merely increasing the blast-wave energy. The steep optical rise indicates that the interaction is dynamically significant, while the faster $\Delta t/t<1$ variability requires structure within the refreshed ejecta or shocked region. If the latter is dominated by angular light-travel-time effects, the measured durations correspond to $\Gamma\theta_{\rm patch}\sim0.48$--$0.62$. Radial ejecta structure or a reverse-shock contribution could instead account for the rapid modulation. Thus, the broad rebrightening, spectral evolution, and fast substructure can be accommodated within a structured refreshed-shock picture without requiring a separate late internal-emission component, although the detailed shock geometry and particle-acceleration physics remain unconstrained.

More generally, GRB~251013C demonstrates the diagnostic value of high-cadence optical monitoring during the early afterglow phase. The ability to resolve both the broad rebrightening and the superposed shorter-timescale structure is essential for separating variability components and for testing whether the emission can be described by a single external-shock process. 

\section{Conclusion}
\label{sec:conclusion}

LAST observations of GRB~251013C reveal pronounced multiscale variability in the early optical afterglow. The light curve shows broad rebrightening episodes with $\Delta t\sim t$, together with resolved shorter-timescale structure during the decay of the main rebrightening, with $\Delta t/t<1$. The main rebrightening also exhibits a structured rise, including a distinct steepening prior to the peak.

The autoregression and fast-rise, exponential-decay analyses provide complementary empirical descriptions of the post-peak substructure and indicate that the deviations from a smooth power-law decay are temporally correlated rather than consistent with independent point-to-point scatter alone. Contemporaneous \textit{Swift}-XRT observations show flaring activity and significant spectral evolution, including a hardening of the photon index during the optical rebrightening. The contemporaneous optical flux remains consistent with an extrapolation of the X-ray spectrum, indicating broadband spectral evolution rather than an isolated change in the X-ray band.

Taken together, these observations favor a structured refreshed-shock interpretation for the main rebrightening. Continued LAST high-cadence observations will provide a larger sample of similarly well-sampled optical afterglows, enabling systematic tests of how common such multiscale variability is in GRB afterglows.
    
\begin{acknowledgements}
R.K. is grateful for the support of the Dean of Faculty fellowship. E.O.O. is grateful for the support of grants from the Willner Family Leadership Institute, André Deloro Institute, Paul and Tina Gardner, The Norman E Alexander Family, M Foundation, ULTRASAT Data Center Fund, Israel Science Foundation, Israeli Ministry of Science, Minerva, BSF, BSF-transformative, NSF-BSF, Israel Council for Higher Education (VATAT), Sagol Weizmann-MIT, Yeda-Sela, and Weizmann-UK. This work made use of data supplied by the UK Swift Science Data Centre at the University of Leicester.
\end{acknowledgements}

\bibliography{refs.bib}

\bibliographystyle{aa}

\begin{appendix}
\nolinenumbers

\section{Field-star photometric control}
\label{app:field_stars}

We tested whether the observed short-timescale variability of the afterglow is an artifact of the observing conditions, the instrument, or the photometric procedure. We did this by extracting the light curves of nearby stars within the same images and analyzing their temporal trends. Source catalogs were cross-matched within $2\arcsec$ against the catalog with the most sources, yielding $449$ sources of which $258$ are detected in at least $95\%$ of the images. We then selected five stars that are within $4\arcmin$ of the burst, are detected in all $635$ images, and cover the brightness range of the afterglow ($15.0 \lesssim m \lesssim 17.4$).

\begin{table}
\centering
\caption{Field stars used as photometric controls.}
\label{tab:field_stars}
\setlength{\tabcolsep}{2.5pt}
\footnotesize
\begin{tabular}{lrrrrrrr}
\hline
Star & R.A. & Dec. & Sep. & $m$ & rms & $\sigma_{\rm phot}$ & $\Delta m$ \\
     & (deg) & (deg) & ($'$) & \multicolumn{4}{c}{(mag)} \\
\hline
209 & 345.83626 & $-0.17043$ & 2.39 & 15.01 & 0.018 & 0.010 & $+0.012$ \\
148 & 345.83504 & $-0.25874$ & 2.91 & 15.84 & 0.028 & 0.015 & $+0.012$ \\
136 & 345.81076 & $-0.27249$ & 4.02 & 16.34 & 0.038 & 0.020 & $+0.030$ \\
137 & 345.82340 & $-0.27176$ & 3.76 & 17.07 & 0.062 & 0.028 & $-0.041$ \\
218 & 345.83465 & $-0.16016$ & 3.01 & 17.36 & 0.083 & 0.033 & $-0.002$ \\
\hline
\end{tabular}
\tablefoot{Separations are from the burst position. $m$ and the rms are computed from the $635$ single-exposure PSF magnitudes, so the rms is a night average. $\sigma_{\rm phot}$ is the median formal uncertainty, and $\Delta m$ the total drift from a linear fit over the full $\sim\!15$~ks interval.}
\end{table}

\begin{figure}
\centering
\includegraphics[width=\columnwidth]{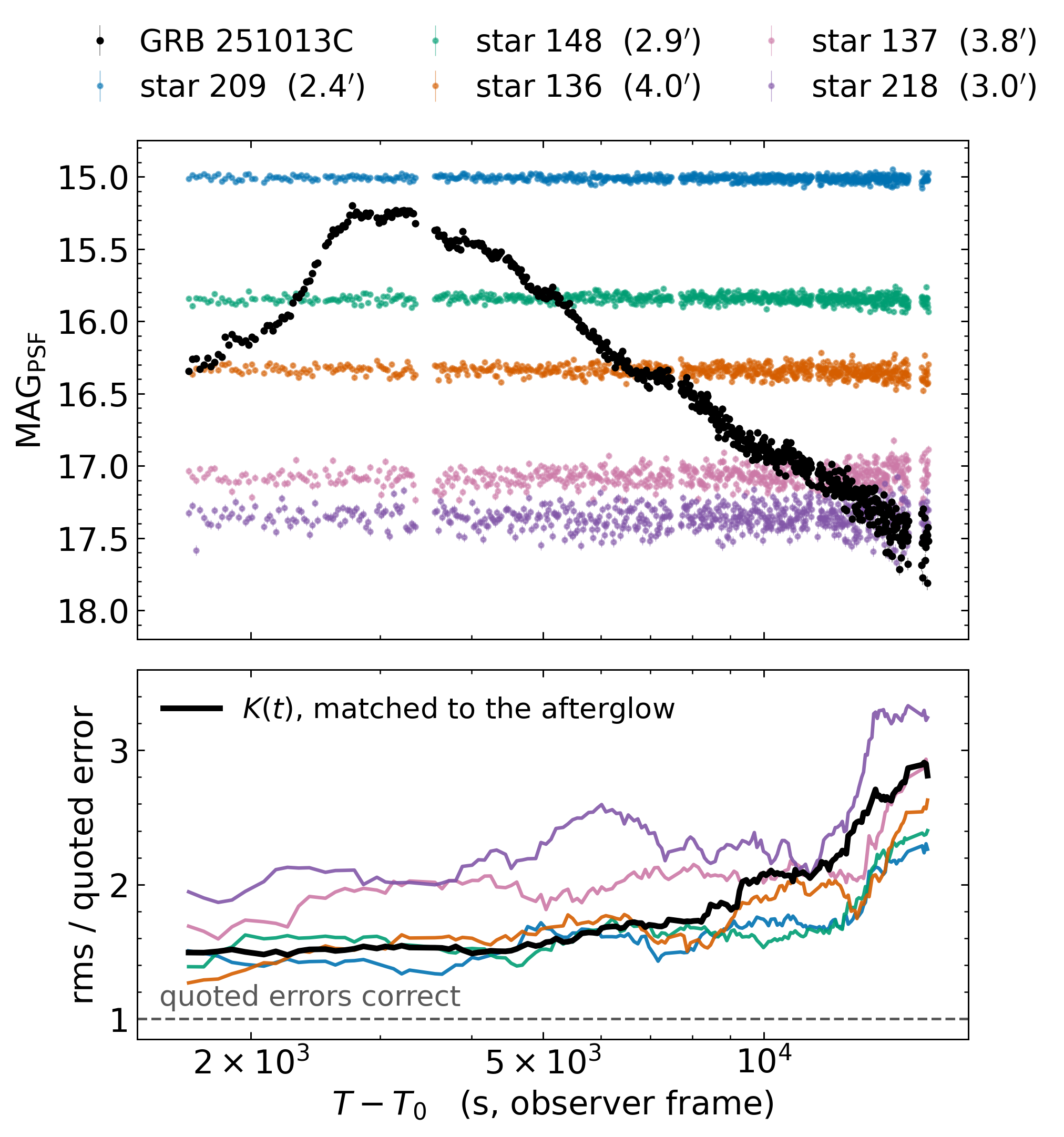}
\caption{Field-star photometric control for the GRB~251013C afterglow. The top panel shows the LAST light curves of the afterglow (black) and of five stars within $4\arcmin$ of the burst. The stars cover the brightness range spanned by the afterglow. None of the stars show similar fluctuations to the afterglow. The bottom panel shows the ratio of measured scatter and the photometric pipeline uncertainties for each of the five stars in the same colors. The curves are produced with a running window of $100$ exposures. The black curve labeled as $K(t)$ is constructed from stars matched to the afterglow in brightness and field position at each epoch.}
\label{fig:field_stars}
\end{figure}

None of the control stars reproduce the structure seen in the afterglow (Tab.~\ref{tab:field_stars}, Fig.~\ref{fig:field_stars}). The scatter of the control stars increases from $0.018$~mag at $m=15.0$ to $0.083$~mag at $m=17.4$, following the photon-noise expectation. Linear fits to the light curves give drifts below $0.003$~mag~ks$^{-1}$. We did however find that across the $258$ well-detected stars, the scatter exceeds the uncertainties produced by the pipeline by a median factor of $2.2$. We further tested for coherency in the scatter and derived a correction factor for the uncertainties of the afterglow.

\subsection{Test for coherent systematics}
\label{app:coherence}

An underestimation of the photometric uncertainties matters only if it is coherent between exposures, since only a coherent component can imitate structure on the timescales of interest. We therefore tested the binning behavior of the field-star light curves by averaging $n$ consecutive exposures. In this case, independent noise is expected to be suppressed as $n^{-1/2}$, whereas a coherent component does not average down and would leave the binned scatter on a plateau above the expected line. Out to the $400$~s visit length, we find that the scatter follows the $n^{-1/2}$ expectation to within $11\%$, falling from $0.0546$~mag in a single exposure to $0.0136$~mag over $20$ exposures, compared with the $0.0122$~mag expected from independent noise alone. Subtracting the independent contribution in quadrature, $\sqrt{0.0136^2-0.0122^2}$, we obtain an upper limit of $0.006$~mag on any coherent component, which is an order of magnitude smaller than the $\sim0.1$~mag fluctuations observed in the afterglow.

Additionally, we tested for a common-mode drift, which would shift all stars together and would therefore not average down. For this purpose, we fitted a relative zero point independently for each exposure using the $258$ well-detected stars. The resulting zero point has an rms of $0.006$~mag and a full range of $0.037$~mag, indicating that the photometric scale carries no coherent trend. The growth of the scatter (Fig.~\ref{fig:field_stars}) is therefore due to a loss of per-exposure precision and not a shift of the zero point. We conclude that neither the photometric procedure nor the observing conditions can account for the fluctuations observed in the afterglow.

\subsection{Calibration of the photometric uncertainties}
\label{app:errcal}

The underestimation of the uncertainties is not uniform. It is lowest near $m= 15.75$, where it falls to a ratio of $1.8$, and rises towards both ends of the magnitude range, reaching $2.5$ for the brightest stars and $2.7$ for the faintest. In addition, the ratio grows through the night from $1.3$--$2.0$ to $2.3$--$3.2$ (Fig.~\ref{fig:field_stars}, lower panel). This behavior appears to track a degradation of the median stellar FWHM during the night from $3.5$ to $4.1$~pixels. Since the control stars hold a constant brightness, we attribute this growth to the observing conditions and not the source flux.

The ratio applicable to the afterglow is the one measured at its own magnitude and field position. Accordingly, at each epoch we took the median ratio over the $16$ to $28$ control stars that lie within $0.75$~mag of the afterglow magnitude and within $10\arcmin$ of its position. The resulting ratio $K(t)$ rises from $1.49$ to $2.81$ across the night, with a median of $1.85$. Over the high-S/N decay interval ($1900 < T-T_0 < 6100$~s), it runs from $1.49$ to $2.03$ with a median of $1.68$, below the all-star value of $2.2$. We use this correction factor in Sect.~\ref{sec:analysis}.

\section{Choice of the baseline model}
\label{app:baseline}

The residual analysis rests on two decisions. The baseline is constrained to lie below the data, and it is fitted to the decay alone. They are not independent, and we set out here the reasoning that leads to both.

\subsection{Definition}
\label{app:baseline_def}

The baseline is a single PL, $F_{\rm PL}(t) = A\,(t/t_{\rm ref})^{\alpha}$ with $t_{\rm ref}=1900$~s and is fitted simultaneously with the AR coefficients. Writing $r_i = F_i - F_{\rm PL}(t_i)$ and $\hat{r}_i = \sum_{k=1}^{p}\beta_k\,r_{i-k}$, the fit minimizes
\begin{align}
S =\ & \sum_{i>p}\left[\frac{r_i-\hat{r}_i}{\sigma_i}\right]^{2} \nonumber\\
     & + w^{2}\sum_i \max\!\left(0,\ \frac{F_{{\rm PL},i}-F_i-\sigma_i}{\sigma_i}\right)^{\!2},
\label{eq:objective}
\end{align}
with $w=50$. The first term is the fit itself, in which the baseline and the AR coefficients are determined together. Fitting the baseline separately by ordinary least squares would instead treat the correlated residual structure as independent scatter and draw the baseline towards the rebrightening episodes. The second term is the envelope constraint discussed in App.~\ref{app:envelope}, which imposes $F_{{\rm PL},i} \le F_i + \sigma_i$ as a penalty rather than as a hard bound. The weight $w$ sets how strongly this penalty acts. The value carries no physical meaning. Any $w$ between $5$ and $200$ leaves the fit unchanged, since over that range the inequality is already satisfied to better than $10^{-2}\sigma$.

\subsection{Why the baseline tracks the lower envelope}
\label{app:envelope}

The residual variability is not small compared with the trend against which it is measured. Part of it evolves slowly enough that a PL can follow it by adjusting its index, and an unconstrained fit does exactly that. Imposing the second term of Eq.~(\ref{eq:objective}) yields $\alpha = -1.25$ and $t_{\rm corr}/t = 0.51$, with $96\%$ of the measurements lying above the baseline. However, if this term is omitted, the same fit steepens to $\alpha = -1.41$ and returns $t_{\rm corr}/t = 0.06$, with the baseline running through the middle of the data and $50\%$ of points above it.

In this case, the steeper index allows the baseline to absorb the slow modulation, which is therefore no longer present in the residual sequence, and the inferred correlation scale collapses by nearly an order of magnitude. In general, any freedom given to the baseline is taken from the residual, so that the direction of the resulting bias is known. We therefore constrain the baseline to the lower envelope, which assigns as little variability to it as possible. The residuals then represent excess emission above a monotonically decaying component and $t_{\rm corr}$ is an upper estimate, which is the conservative direction for the result we claim.

\subsection{Why the fit is restricted to the decay}
\label{app:decayonly}

\begin{table}
\centering
\caption{Baselines spanning the full broad flare.}
\label{tab:baselines}
\setlength{\tabcolsep}{3pt}
\footnotesize
\begin{tabular}{lcc}
\hline
Baseline & $\chi^2_\nu$ & pre-kink \\
\hline
PL $+$ broken-PL        & 1.579 & 0.25 \\
Two breaks              & 1.564 & 0.29 \\
Two breaks, anchored    & 1.546 & 0.02 \\
\hline
\end{tabular}
\tablefoot{All three are fitted over $1000 < T-T_0 < 11000$~s with the objective of Eq.~(\ref{eq:objective}), the last additionally carrying a term penalizing the baseline for lying below the data. The pre-kink column is the median fractional offset between the data and the baseline over $1040 < T-T_0 < 1440$~s, so a baseline that follows the shallow section before the kink brings it near zero. Values of $\chi^2_\nu$ are comparable between rows but not with the fiducial fit, which spans a different interval.}
\end{table}

\begin{figure}
\centering
\includegraphics[width=\columnwidth]{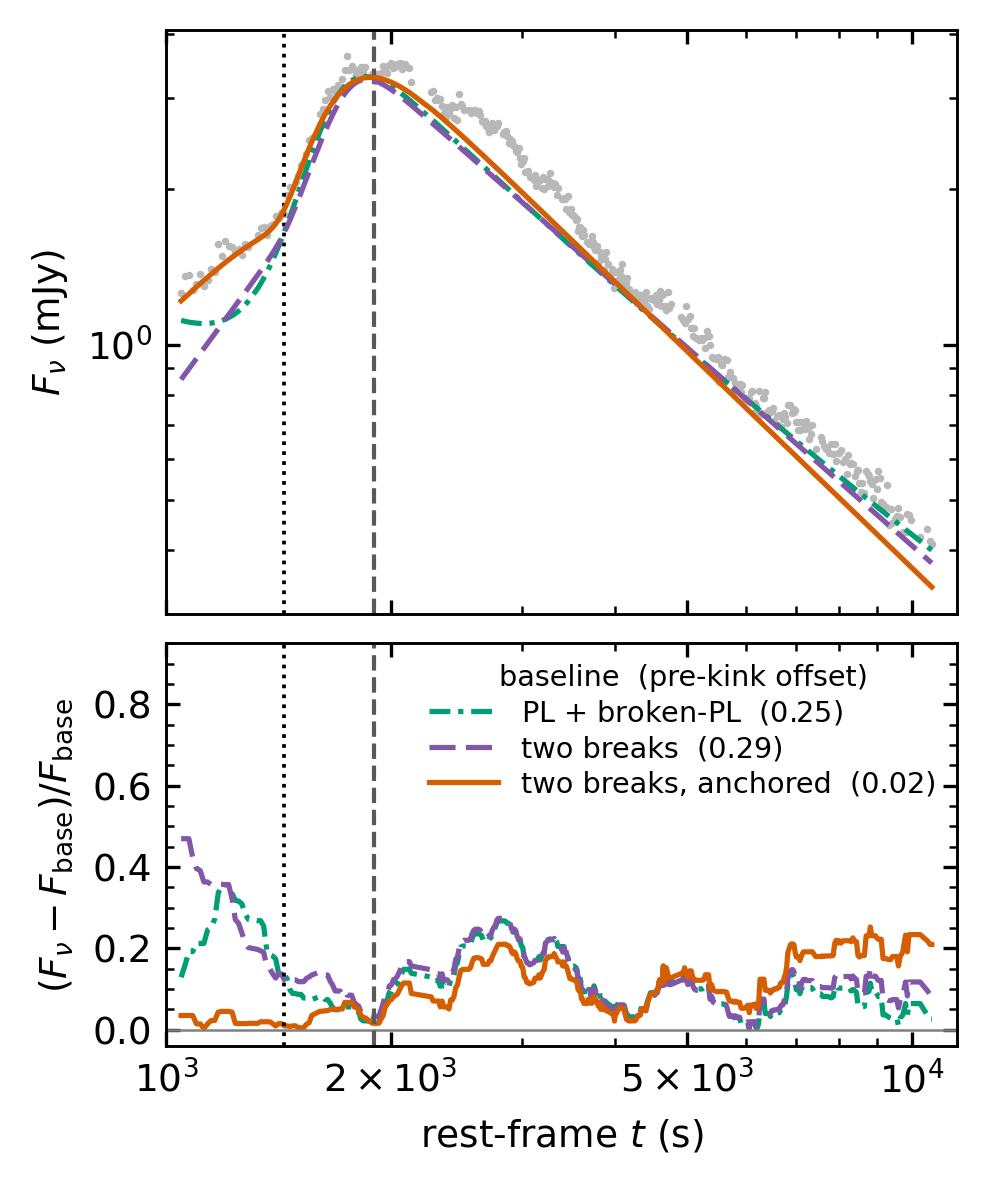}
\caption{Baselines spanning the full broad flare. The top panel shows the rebinned light curve with the three baselines of Tab.~\ref{tab:baselines}, labeled with the pre-kink residual defined there. The bottom panel shows the fractional residual about each, smoothed with a running median over $7$ points, about $90$~s at the median cadence. The dotted line marks the kink in the rise at $1440$~s and the dashed line the start of the fiducial interval at $1900$~s. The residuals trace common structure after the fiducial cut and diverge before it.}
\label{fig:baselines}
\end{figure}

Short-timescale structure is present during the rise as well as the decay, so we tested baselines describing the full broad flare, fitted over $1000 < T-T_0 < 11000$~s with the same objective. We retained the envelope constraint here, since the decay still dominates. Three baselines were kept (Tab.~\ref{tab:baselines}, Fig.~\ref{fig:baselines}): (i) an underlying PL plus an additive smoothly broken PL; (ii) a chained doubly broken PL in which a break is free to fall near the kink at $\sim\!1440$~s; and (iii) the same doubly broken PL as (ii) but anchored by a further term penalizing the baseline for lying below the data. A PL plus a FRED was discarded, since a FRED turn-on is smooth by construction and cannot reproduce the kink.

Over the decay the three agree to a median of $4.2\%$, with a maximum difference of $16\%$, the largest departures coming from the anchored model, which is held above the others on the rise and compensates with a steeper decay. Over the rise, however, only the anchored model follows the shallow section preceding the kink, the other two lying $25\%$ and $29\%$ below the data there. The anchored model is itself poorly determined. Since the coverage begins part-way through the rise at $1050$~s, only $30$ of the $397$ measurements precede the kink. Between the kink and the peak, the three baselines agree to $5\%$, but over those $30$ points they differ by $32\%$, and the pre-kink index is not constrained. Fixing this index on a grid and refitting the remaining parameters changes the objective by only $1.7\%$ over the full index range of $0.05$--$2$ and leaves no well-defined minimum within it. These fits are therefore limited by the imposed parameter ranges and not by the data, coming to rest on a new limit each time one is widened.

The decay is therefore the only interval where the broad component and the superposed variability separate reliably, and it carries most of the flare. We fitted it alone, and we treat the rise as qualitative evidence that the variability extends to earlier times.

\section{Robustness of the autoregressive decorrelation scale}
\label{app:ar_robustness}

We tested the robustness of the AR-derived decorrelation scale under changes to the AR order, the temporal window, the temporal ordering of the residuals, and the sampling cadence. The fiducial constrained PL+AR fit gives
\begin{equation}
    t_{\rm corr}/t \simeq 0.51 ,
\end{equation}
where \(t_{\rm corr}\) is defined as the first lag at which the absolute AR-implied autocorrelation function falls below \(1/e\).

\begin{figure}
\centering
\includegraphics[width=\columnwidth]{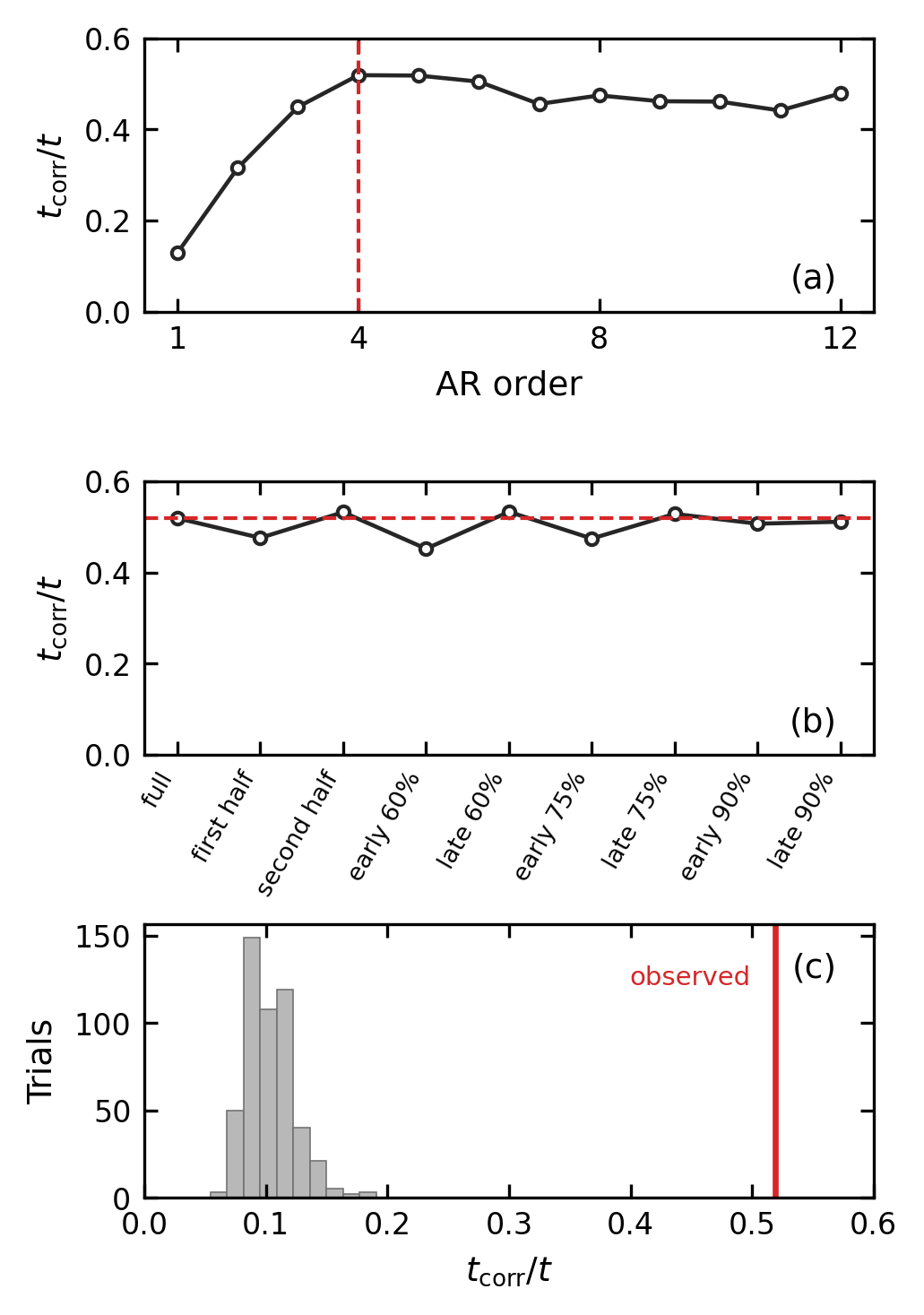}
\caption{Robustness of the decorrelation scale. Panel (a): $t_{\rm corr}/t$ as a function of the adopted AR order, with the red dashed line marking the selected order. Panel (b): $t_{\rm corr}/t$ obtained from the fixed fiducial AR-implied ACF while varying only the observation-time window used to map its sample-lag decorrelation scale into $\Delta t/t$, with the red dashed line marking the full-window value. Panel (c): distribution of $t_{\rm corr}/t$ over $500$ realizations in which the fitted residuals are randomly permuted, removing their temporal ordering, with the red line marking the observed value. All three panels share the same range in $t_{\rm corr}/t$.}
\label{fig:GRB251013C_appendix_ar_robustness}
\end{figure}

First, we repeated the analysis for different AR orders across the reduced-\(\chi^2\) plateau (Fig.~\ref{fig:GRB251013C_AR_order_delta_chi2}). The inferred decorrelation scales (Fig.~\ref{fig:GRB251013C_appendix_ar_robustness}a) have
\begin{equation}
    {\rm median}(t_{\rm corr}/t)=0.46,
\end{equation}
with a 16--84 percentile range of
\begin{equation}
    t_{\rm corr}/t = 0.41\text{--}0.50.
\end{equation}
The selected AR(4) model therefore lies toward the longer-timescale end of the plateau and does not correspond to an anomalously short decorrelation scale.

We also recomputed the normalized decorrelation scale in several temporal sub-windows of the fitted interval (Fig.~\ref{fig:GRB251013C_appendix_ar_robustness}b). The inferred scale is stable across these choices. The sub-windows span $t_{\rm corr}/t\simeq0.45$--$0.53$. The early windows give slightly smaller values, $t_{\rm corr}/t\simeq0.45$--$0.50$, and the late windows give $t_{\rm corr}/t\simeq0.50$--$0.53$, but this variation does not alter the main conclusion. 

To test whether the measured correlation scale could arise from the residual amplitude distribution alone, we randomly shuffled the residual sequence and repeated the AR analysis (Fig.~\ref{fig:GRB251013C_appendix_ar_robustness}c). The shuffled realizations give
\begin{equation}
    {\rm median}(t_{\rm corr}/t)=0.10,
\end{equation}
with a 5--95 percentile range of
\begin{equation}
    t_{\rm corr}/t = 0.07\text{--}0.13 .
\end{equation}
The fiducial value lies above this range, indicating that the inferred correlation scale depends on the temporal ordering of the residuals.

Adaptive rebinning produces a time-dependent sampling cadence, with broader bins at late times. We therefore tested this dependence by restricting the fit to the high-S/N interval, $1900~\mathrm{s}< T-T_0<6100~\mathrm{s}$. The mean rest-frame cadence is $26.1$~s over the full window and $18.0$~s over the restricted high-S/N window. The preferred order changes from AR(4) to AR(3), but the inferred decorrelation timescale remains similar at $0.48$, indicating that the inferred physical timescale is robust to the changing sampling cadence.

Finally, we generated noise-only realizations using the same sampling and photometric uncertainties. These decorrelate on the sampling scale, with
\begin{equation}
    {\rm median}(t_{\rm corr}/t)=0.0044 .
\end{equation}
Thus, uncorrelated photometric noise does not reproduce the observed AR decorrelation scale.

\end{appendix}
\end{document}